\documentclass[useAMS,fleqn, usenatbib, final]{mnras}

\usepackage{aas_macros}
\usepackage{amssymb}
\usepackage{widebar}
\usepackage{ragged2e}
\usepackage{amsmath}
\usepackage{array}
\usepackage{units}
\usepackage[dvipsnames]{xcolor} 
\usepackage{lineno}
\usepackage{dcolumn}
\usepackage[normalem]{ulem} 
\usepackage{subfigure}
\usepackage{xspace} 
\usepackage{bm}
\usepackage{graphicx} 
\usepackage[bottom]{footmisc}
\usepackage{hyperref}
\usepackage{color,units}
\usepackage{aas_macros}
\usepackage[T1]{fontenc}
\usepackage{ae,aecompl}
\usepackage{newtxtext,newtxmath}

\usepackage{longtable}
\usepackage{booktabs}
\usepackage{multirow}
\usepackage{colortbl}
\usepackage{nicematrix}

\graphicspath{{images/}} 

\newcommand{\code}[1]{{\texttt{#1}}\xspace}
\newcommand{\bilby}{\code{bilby}}
\newcommand{\bilbypipe}{\code{bilby-pipe}}
\newcommand{\dynesty}{\code{dynesty}}
\newcommand{\gwpy}{\code{GWpy}}

\newcommand{\imrphenomp}{{\sc \texttt{IMRPhenomPv2}}\xspace}
\newcommand{\seob}{{\sc \texttt{SEOBNRv4PHM}}\xspace}
\newcommand{\nrsur}{{\sc \texttt{NRSur7dq4}}\xspace}
\newcommand{\imrxhm}{{\sc \texttt{IMRPhenomXPHM}}\xspace}
\newcommand{\gstlal}{{\sc GstLAL}\xspace}
\newcommand{\cwb}{{\sc cWB}\xspace}
\newcommand{\spiir}{{\sc SPIIR}\xspace}

\newcommand{\pycbc}{{\sc {{PyCBC}}}\xspace}
\newcommand{\GWTC}{{\sc {{GWTC-1}}}\xspace}

\newcommand{\IAS}{{\sc {{IAS}}}\xspace}

\newcommand{\fancytext}[1]{{\relax\ifmmode#1\else $#1$\fi}\xspace}
\newcommand{\mathcmd}[1]{{\sc \relax\ifmmode#1\else $#1$\fi}\xspace}
\newcommand{\bcr}{\mathcmd{\rho_\text{BCR}}}
\newcommand{\bgrdbcr}{\mathcmd{\rho_\text{BCR}^\textit{b}}}
\newcommand{\candbcr}{\mathcmd{\rho_\text{BCR}^\textit{c}}}
\newcommand{\injbcr}{\mathcmd{\rho_\text{BCR}^\textit{s}}}

\newcommand{\psd}{\mathcmd{P(f)}}
\newcommand{\msun}{\mathcmd{M_\odot}}

\newcommand{\parameters}{\mathcmd{\vec{\theta}}}

\newcommand{\template}{\mathcmd{\mu(\parameters)}}
\newcommand{\fap}{\mathcmd{p_{\textit{B}}}}
\newcommand{\totMlab}{\mathcmd{M}}
\newcommand{\pastro}{\fancytext{p_\text{astro}}}

\newcommand{\pastrobcr}{\fancytext{p_{\widebar{B}}}}
\newcommand{\untunedpastrobcr}{\fancytext{p_{\widebar{B}}^{\prime}}}
\newcommand{\tunedpastrobcr}{\fancytext{p_{\widebar{B}}}}

\newcommand{\pastroGwtcPycbc}{\fancytext{p_\text{astro}^{\text{pyCBC}}}}
\newcommand{\pastroGwtcGstlal}{\fancytext{p_\text{astro}^{\text{GstLAL}}}}
\newcommand{\pastroIas}{\fancytext{p_\text{astro}^{\text{IAS}}}}
\newcommand{\pastroPrat}{\fancytext{P(S|\text{d})}}

\newcommand{\pastroOgcTwo}{\fancytext{p_\text{astro}^{\text{OGC2}}}}
\newcommand{\pastroOgcThree}{\fancytext{p_\text{astro}^{\text{OGC3}}}}
\newcommand{\tc}{\fancytext{t_c}} 

\renewcommand{\aboverulesep}{0pt}
\renewcommand{\belowrulesep}{0pt}
\newcolumntype{C}{>{\centering\arraybackslash}X}
\newcolumntype{L}{>{\arraybackslash}X}
\newcolumntype{a}{>{\columncolor{gray}}c}
\newcommand{\SPA}{School of Physics and Astronomy, Monash University, Clayton VIC 3800, Australia}
\newcommand{\OzGravMonash}{OzGrav: The ARC Centre of Excellence for Gravitational Wave Discovery, Clayton VIC 3800, Australia}
\newcommand{\CIT}{LIGO Laboratory, California Institute of Technology, Pasadena, CA 91125, USA}
\newcommand{\MIT}{LIGO Laboratory, Massachusetts Institute of Technology, Cambridge, MA 02139, USA}
\newcommand{\Kavli}{Department of Physics and Kavli Institute for Astrophysics and Space Research,\\ Massachusetts Institute of Technology, 77 Massachusetts Ave, Cambridge, MA 02139, USA}
\newcommand{\Royal}{Department of Physics, Royal Holloway, University of London, TW20 0EX, United Kingdom}

\title[An IMBH candidate follow-up in O2 using the BCR]{A follow-up on intermediate-mass black hole candidates in the second LIGO--Virgo observing run with the Bayes Coherence Ratio}

\author[Vajpeyi et al.]{Avi~Vajpeyi$^{1,2}$\thanks{avi.vajpeyi@monash.edu},
    Rory~Smith$^{1,2}$,
    Eric~Thrane$^{1,2}$,
    Gregory~Ashton$^{1,2,3}$,
    Thomas~Alford$^{4}$,
    \newauthor
    Sierra~Garza$^{4}$,
    Maximiliano~Isi$^{5,6}$,
    Jonah~Kanner$^{4}$,
    T. J.~Massinger$^{4}$,
    Liting~Xiao$^{4}$
\\
$^{1}$ \SPA \\
$^{2}$ \OzGravMonash \\
$^{3}$ \Royal \\
$^{4}$ \CIT \\
$^{5}$ \MIT \\
$^{6}$ \Kavli 
}

\date{Last updated XXX; in original form XXX}

\pubyear{2021}

\begin{document}
\label{firstpage}
\pagerange{\pageref{firstpage}--\pageref{lastpage}}
\maketitle

\begin{abstract}
The detection of an intermediate-mass black hole population ($10^2-10^6\ M_\odot$) will provide clues to their formation environments (e.g., disks of active galactic nuclei, globular clusters) and illuminate a potential pathway to produce supermassive black holes. 
Ground-based gravitational-wave detectors are sensitive to mergers that can form intermediate-mass black holes weighing up to $\sim 450\ M_\odot$.
However, ground-based detector data contain numerous incoherent short duration noise transients that can mimic the gravitational-wave signals from merging intermediate-mass black holes, limiting the sensitivity of searches. 
Here we follow-up on binary black hole merger candidates using a ranking statistic that measures the coherence or incoherence of triggers in multiple-detector data. 
We use this statistic to rank candidate events,  initially identified by all-sky search pipelines, with lab-frame total masses $\gtrsim55\ M_\odot$ using data from LIGO's second observing run. 
Our analysis does not yield evidence for new intermediate-mass black holes. 
However, we find support for eight stellar-mass binary black holes not reported in the first LIGO-Virgo gravitational wave transient catalog GWTC-1, seven of which have been previously reported by other catalogs.
\end{abstract}


\section{Introduction}
\label{sec:intro}

Stellar mass (${M_\text{BH} < 10^{2}~\msun}$) and supermassive black holes (${M_\text{BH} > 10^{6}~\msun}$) have been observed and well studied since the 1970s~\citep{Webster:1972:Natur, Balick:1974:ApJ, Ghez:1998:ApJ, Genzel:2010:RvMP, Abbott:2019:PhRvX, EventHorizonTelescopeCollaboration:2019:ApJL, Abbott:2020:arXiv}. 
However, there is a deficiency of observational evidence for black holes in the intermediate-mass range ${10^{2} - 10^{6}~\msun}$. 
A variety of techniques have been employed to search for intermediate-mass black hole (IMBH) candidates including reverberation mapping~\citep{Peterson:2014:SSRv}, direct kinematic measurements~\citep{Schodel:2002:Natur, Kiziltan:2017:Natur}, applying macroscopic galaxy to black hole mass scaling relations, $M_{BH}$-$\sigma$ and $M_{BH}$-L relations~\citep{Graham:2013:ApJ, Wevers:2017:MNRAS}, studying  X-ray luminosity and spectra~\citep{Greene:2004:ApJ, Lin:2020:ApJL}, gravitational lensing of gamma-ray burst light curves~\citep{paynter_evidence_2021}, and others~\citep[see][]{Greene:2020:ARA&A, Koliopanos:2017:mbhe, Mezcua:2017:IJMPD}.
However, because IMBH have smaller masses than those of supermassive black holes, it is much more challenging to observe them with these observational techniques~\citep{Mezcua:2017:IJMPD}. 
Additionally, numerous IMBH candidates discovered using these techniques are ambiguous as the observations can be attributed to other sources (e.g., light sources orbiting clusters of stellar-mass black holes~\citealt{Ridolfi:2016:MNRAS, Freire:2017:MNRAS},  anisotropic emission from neutron stars~\citealt{Israel:2017:MNRAS, RodriguezCastillo:2020:ApJ}).
The discovery of an IMBH population will bridge the intermediate-mass observational gap, probe IMBH formation environments (e.g. accretion disks of active galactic nuclei~\citealt{Tagawa:2021:ApJ, Li:2021:arXiv, Samsing:2020:arXiv, Tagawa:2020:ApJ, Ishibashi:2020:A&A, Grobner:2020:A&A, Yang:2019:PhRvL, McKernan:2019:ApJL, Yang:2019:ApJ, McKernan:2018:ApJ, Bellovary:2016:ApJL, McKernan:2014:MNRAS, McKernan:2012:MNRAS}, the centers of dense stellar clusters~\citealt{Banerjee:2021:MNRASa, Zevin:2021:ApJ,Mapelli:2021:arXiv,Weatherford:2021:ApJL, Bouffanais:2021:arXiv, Ballone:2021:MNRAS, Kumamoto:2021:arXiv, Banerjee:2021:MNRASb, Martinez:2020:ApJ, Romero-Shaw:2020:ApJL, Anagnostou:2020:PASA}, Population-III stars~\citealt{Toubiana:2021:PhRvL, Farrell:2021:MNRAS, Safarzadeh:2020:ApJL, Liu:2020:MNRAS, Inayoshi:2017:MNRAS}), and illuminate our understanding of supermassive black hole formation~\citep{Askar:2021:MNRAS, ArcaSedda:2019:arXiv, Amaro-Seoane:2007:CQGra, Gurkan:2006:ApJL}. 

Compact binary coalescences (CBCs) can provide gravitational-wave signals for IMBH candidates e.g., the ${142^{+28}_{-16}~\msun}$ ($90\%$ credible intervals) remnant observed from the gravitational-wave event GW190521~\citep{Abbott:2020:PhRvL} and other candidates~\citep{ligo_imbh_search, ligo_imbh_o3, pycbc_imbh}. As a binary's lab-frame total mass $M$ is associated with its gravitational-wave merger frequency, ${f\propto M^{-1}}$,  ground-based gravitational-wave detectors (${f\sim 10^1 - 10^3\ \text{Hz}}$) are sensitive to the last milliseconds of merging systems with ${100~\msun < \totMlab < 400~\msun}$~\citep{LIGOScientificCollaboration:2015:CQGra, Martynov:2016:PhRvD, Moore_2014, Acernese:2015:CQGra}, while space-based detectors (${f\sim 10^{-2}-10^1\ \text{Hz}}$) can study the full signals of merging systems with ${10^4~\msun < \totMlab < 10^7~\msun}$~\citep{ Moore_2014, Lu:2019:PhRvD}. Because of the short duration of IMBH gravitational-wave signals in ground-based detectors, data quality is critical for their detection. Gravitational-wave data is characterized by numerous non-stationary terrestrial artifacts called \textit{glitches}~\citep{ pycbc_short_duration_transients, pe_with_glitch, blip_glitches}. Like signals from IMBH mergers, most glitches last for a fraction of a second, making them difficult to distinguish from astrophysical signals. These glitches can decrease the sensitivity of searches for binary black hole mergers with ${\totMlab\gtrsim55~\msun}$ ~\citep{pycbc_short_duration_transients}.

Although a significant fraction of the glitches can be identified by testing them for coherence amongst two or more detectors and performing matched-filtering, these methods are insufficient to identify all glitches~\citep{ pycbc_short_duration_transients, pe_with_glitch, blip_glitches}. One method to discriminate more glitches while searching for CBCs is the Bayesian odds~\citep{bci, kanner2016leveraging, BCR1, BCR2, bcr_gw151216, bayesian_odds}. The Bayesian Coherence Ratio \bcr~\citep{BCR1,BCR2} is a Bayesian odds comparing the probability that the data contains coherent signals vs. incoherent glitches. In this paper, we use the \bcr to rank O2’s coincident CBC gravitational-wave candidates with lab-frame total masses in the range of ${55-500~\msun}$. 
We present the candidates' $\pastrobcr$, the probability that the candidate is inconsistent with the background distributions of \bcr values computed from time-slid data. 
Additionally, for comparison, we provide the candidate's \pastro values reported by the LIGO-Virgo-KAGRA (LVK) collaboration in \GWTC~\citep{GWTC1}, the \pycbc-team~\citep{pycbc_code, pycbc_og0, pycbc_og1, pycbc_og2, pycbc_og3, pycbc_og4, pycbc_og5, pycbc_og6, pycbc_single_det, pycbc_ogc_2}, by the Institute of Advanced study's team (\IAS)~\citep{IAS0, IAS1, IAS2}, and by \citet{bayesian_odds}. 

We find that (a) events reported in \GWTC, including GW170729 (likely the most massive BBH system in \GWTC) are statistically significant ${\pastrobcr>0.9}$; (b) three out of the eight \IAS events and candidates have ${\pastrobcr>0.9}$, corroborating \IAS's detection claims for GW170304, GW170727, and GW170817A; and that (c) our ranking statistic does not identify any new IMBH, but does identify an unreported marginal stellar-mass binary black hole candidate, 170222 with ${\pastrobcr\sim0.6}$.\footnote{
\RaggedRight
170222 is a sub-threshold candidate detected by \pycbc (${{\rm SNR}\sim7.7}$). The prefix of GW is not utilized as this is a candidate event.
} 

The remainder of this paper is structured as follows. We outline our methods, including details of our ranking statistic and the retrieval of our candidates in Section~\ref{sec:method}. We present details on the implementation of our analysis in Section~\ref{sec:Analysis}. Finally, we present our results in Section~\ref{sec:Results} and discuss these results in the context of the significance of gravitational-wave candidates in Section~\ref{sec:Conclusion}.

\section{Method}
\label{sec:method}
\subsection{A Bayesian Ranking Statistic}
The standard framework to identify CBC gravitational-wave signals in data is to quantify the significance of candidates with null-hypothesis significance testing~\citep{GWTC1, GWTC2}. In this framework, the candidates' ranking statistic is compared against a background distribution. The independent matched-filter searches, e.g., \pycbc~\citep{pycbc_og4}, \spiir~\citep{spiir} and \gstlal~\citep{sachdev2019gstlal}, and Coherent WaveBurst~\citep{cwb} used by LVK to search for signals in gravitational-wave data all use ranking statistics in such a manner~\citep{GWTC1}. Both \pycbc and \gstlal's ranking statistic incorporate information about the relative likelihood that the data contains a coherent signal versus noise. In contrast, \cwb's ranking statistic uses the information of coherent energy present in the network of detectors~\citep{GWTC1}. 

Bayesian inference offers an alternative means to rank the significance of candidate events by computing the odds that the data contain a transient gravitational-wave signal versus instrumental glitches~\citep{BCR1}. This method relies on accurate models for the signal and glitch morphologies~\citep{BCR1}. In principle, Bayesian odds is the optimal method for hypothesis testing~\citep{BCR2}. Much of its power comes from the Bayesian evidence, the likelihood of the data given a hypothesis. However, the evidence is not used in current matched filter searches. Here, we explore a hybrid frequentist/Bayesian ranking statistic that makes use of the Bayesian evidence. We compute the Bayesian evidence under the assumption that the data either contain a coherent gravitational-wave signal, noise, or a glitch ($Z^S, Z^N, Z^G$, defined in Appendix~\ref{sec:bayesianEvidEval}). However, because we do not have at our disposal a set of PyCBC triggers generated for simulated signals from a realistic population, we use the evidences as a ranking statistic, instead of computing true Bayesian odds. We form a bootstrapped distribution of the evidence for simulated foreground and background events to form a frequentist ranking statistic.
Our work highlights the importance of an astrophysically realistic injection set for calculating $p_\text{astro}$.

\subsection{Formalism}

Introduced by \citet{BCR1}, the Bayesian Coherence Ratio for a signal in a network of $D$ detectors is given by
\begin{equation}
\label{eq:bcr}
\bcr = \frac{\hat{\pi}^S Z^S}{\prod\limits^D_{i=1} \ [\hat{\pi}^G_i Z^G_i + \hat{\pi}^N_i Z^N_i]}\  ,
\end{equation}
where $\{\hat{\pi}^S, \hat{\pi}^N_i, \hat{\pi}^G_i\}$ are ``pseudo prior probabilities'' that the data contain a coherent signal, incoherent noise or an incoherent glitch. These factors are not true prior probabilities because they are not chosen a priori. 
Rather, these factors are obtained by minimizing the overlap between a signal and background distribution (see Appendix D). 
We assume each detector has the same glitch and noise prior probabilities of $\{\hat{\pi}^N, \hat{\pi}^G\}$. In the limit where our pseudo prior probabilities equal the actual prior probabilities, the \bcr becomes the optimal Bayesian odds described by \citet{BCR2}.  However, as we do not (in this work) have a reliable estimate for the prior probabilities, we cannot interpret the \bcr as a Bayesian odds to discriminate signals from glitches. 
Instead, we use the \bcr as a ranking statistic to obtain a frequentist significance of \bcr.  


Since it is impossible to shield ground-based gravitational-wave detectors from gravitational-wave signals, the LVK empirically estimates the background by repeatedly time-shifting strain data by amounts larger than the light-travel time between the two LIGO detectors~\citep{GWTC1}. We use time-shifted data to generate $\bgrdbcr$, the background ranking statistic. Following this, we calculate the fraction of $\bgrdbcr$ greater than or equal to a $\candbcr$, the candidate ranking statistic:
\begin{equation}
    p_1^b = \frac{\text{Count of } \bgrdbcr \leq \candbcr}{\text{Count of } \bgrdbcr} \ .
\end{equation}
Given a set of simulated signals and their ranking statistic $\injbcr$, one may calculate the fraction of $\injbcr$ greater than or equal to a $\candbcr$: 
\begin{equation}
    p_1^s = \frac{\text{Count of } \injbcr \leq \candbcr}{\text{Count of } \injbcr} \ .
\end{equation}
With $p_1^b$ and $p_1^s$ it is possible to compute a candidate’s 
\pastro, the probability that a candidate is of astrophysical origin:
\begin{equation}
    \pastro = \frac{p_1^s}{p_1^s + p_1^b}\ .
\end{equation}
However, for this study we do not have an astrophysical distribution of simulated signals and so we cannot compute $p_1^s$ or consequently $\pastro$. 
Instead we opt for a frequentist $p$-value probability that a candidate is inconsistent with the background. 
As we have $k$ candidates, each with a \candbcr, we calculate a false-alarm probability \fap that accounts for trial factors given by
\begin{equation}
    \fap = 1 - (1-p_1^b)^k \ .
\end{equation}
Finally, we compute the probability that a candidate is inconsistent with the background:
\begin{equation}
\pastrobcr = 1 - \fap \ . 
\end{equation}
When $\pastrobcr \ll 1$, the event is consistent with the background distribution. 
Conversely, when $\pastrobcr \approx 1$ the event is inconsistent with the background distribution, and is therefore a promising gravitational-wave candidate.

It is important to note that 
$\pastrobcr$ (the probability that an event is not part of the background distribution) is not the same as $\pastro$, which requires an astrophysical set of simulated signals.

\section{Analysis}
\label{sec:Analysis}

We acquire candidate signal triggers (times when the detector's data has a signal-to-noise ratio above a predetermined threshold) for \bcr analysis from \pycbc's search in O2~\citep{pycbc_code, pycbc_og0, pycbc_og1, pycbc_og2, pycbc_og3, pycbc_og4, pycbc_og5, pycbc_og6, public_ligo_o2_triggers}.  Some of the triggers are associated with gravitational-wave events and candidates while others are glitches. We also acquire background time-slid triggers and simulated triggers from \pycbc's O2 search to calculate $\bgrdbcr$ and estimate values for $\{\hat{\pi}^S,\hat{\pi}^G\}$ (see Appendix~\ref{apdx:tuning-prior-odds} for details on the estimation process). The triggers are divided into two week time-frames because the detector's sensitivity does not stay constant throughout the eight-month-long observing period~\citep{pycbc_og4}.

\begin{table}
\caption[BBH parameters correspond to duration $\leq454\ \text{ms}$]{\label{tab:parameters}Trigger-selection lab-frame parameter space (parameters correspond to signals with durations $\leq454 \ \text{ms}$ and $q\geq0.1$).}
\centering
\begin{tabular}{lrr}
\toprule
           & Minimum & Maximum\\
\midrule
Component Mass 1, $m_1$ [\msun] & 31.54 & 491.68\\
Component Mass 2, $m_2$ [\msun] & 1.32 & 121.01\\
Total Mass, $M$ [\msun] & 56.93 & 496.72\\
Chirp Mass, $\mathcal{M}$ [\msun] & 8.00 & 174.56\\
Mass Ratio, $q$ & 0.1 & 0.98\\
\end{tabular}
\end{table}

\begin{figure*}
{\centering \includegraphics[width=0.75\linewidth]{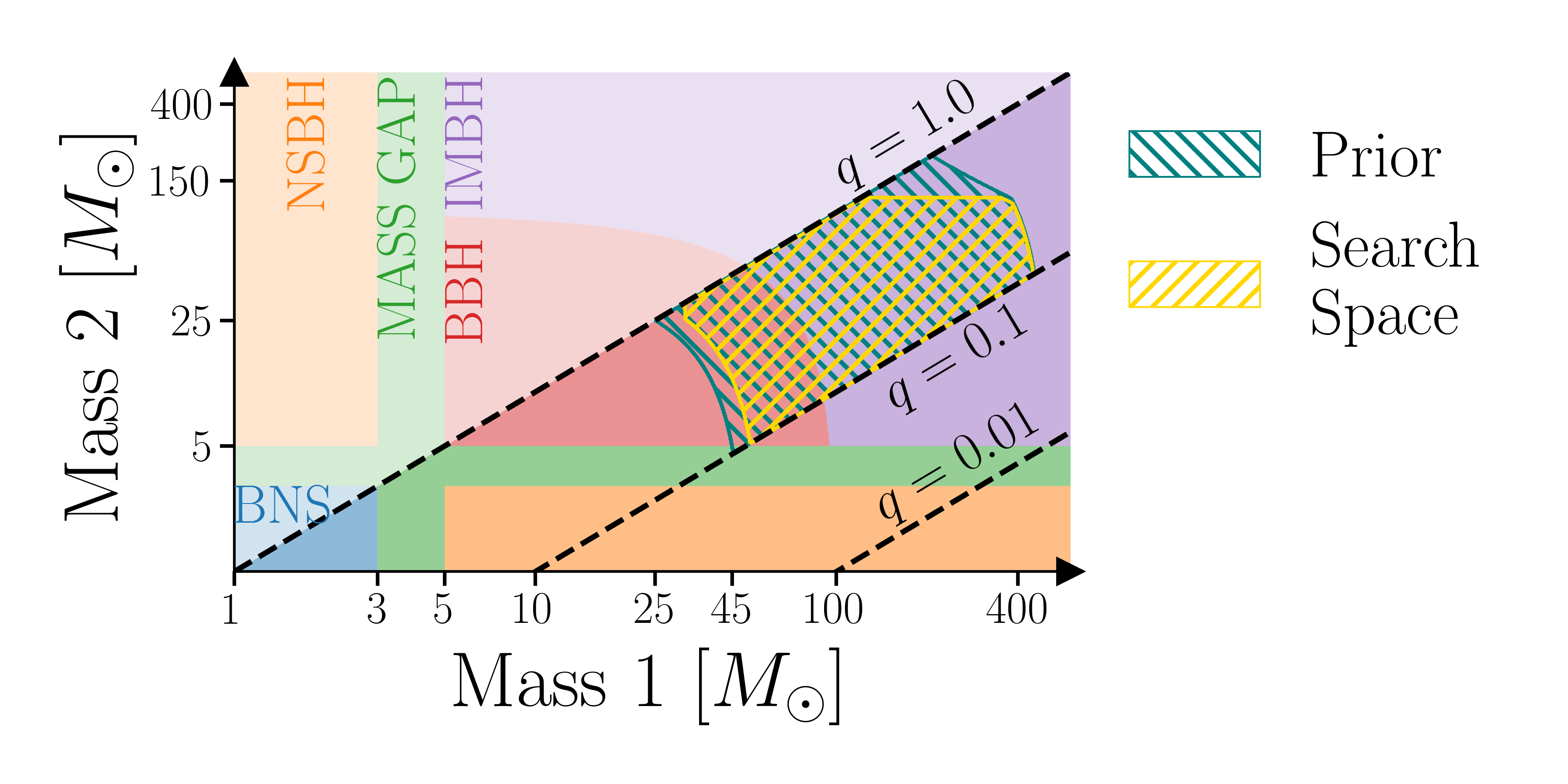}
}
\caption[ BCR search space.]{Lab-frame black hole component-mass boundaries for our search space and parameter estimation prior. Our search is constrained to the parameter space enclosed by the gold-colored hatches, while our prior is constrained to the slightly larger parameter space enclosed by the teal-colored hatches. The purple region labeled ``IMBH'' is the parameter space where merger remnants may be IMBHs.}\label{fig:templateBank}
\end{figure*}

For our study, we filter \pycbc triggers to include only those in the parameter ranges presented in Table~\ref{tab:parameters}. This region focuses our analysis on binary black hole mergers with lab-frame total masses above ${\gtrsim55 \msun}$, corresponding to binary systems with signal durations $<454 \ \text{ms}$ and ${q\geq0.1}$. The filtering process leaves us with ${{\sim}70{,}000}$ background, ${\sim}5{,}000$ simulated, and $25$ candidate signal triggers. We additionally include events and candidate events reported by \GWTC and the \IAS group in our list of candidate signal triggers.  A plot of the lab-fame component mass space constrained by our search space is presented in Fig.~\ref{fig:templateBank}.

To evaluate $\{Z^S, Z^G_i, Z^N_i\}$ and calculate the \bcr (Eq.~\ref{eq:bcr}) for triggers, we carry out Bayesian inference with \bilby~\citep{bilby, bilby_pipe}, employing \dynesty~\citep{dynesty} as our nested sampler. Nested sampling, an algorithm introduced by~\citet{skilling2004, skilling2006}, provides an estimate of the Bayesian evidence and is often utilized for parameter estimation within the LIGO collaboration~\citep{bilby, bilby_paper, pbilby_paper}.

We use a likelihood that marginalizes over coalescence time, the phase at coalescence, and luminosity distance~\citep[see][Eq.~80]{intro_to_gw_bayes}. We use identical parameter estimation priors for the glitch and signal models. We restrict the spin priors to aligned spins to reduce the number of parameters we sample. We define our mass priors to be uniform in chirp mass $\mathcal{M}$ and mass ratio $q$ to avoid sampling issues that arise from sampling in thin regions of the component mass parameter space~\citep{bilby_gwtc}. As a post-processing step, we convert posterior samples calculated with uniform $\{\mathcal{M}, q\}$ priors to uniform component mass priors by re-sampling the posterior samples using the Jacobian given in~\citet[Eq.~21]{Veitch:2015:PhRvD}. The complete list of the priors is in Table~\ref{tab:priors}.

\begin{table}
    \centering
    \caption{
    Prior settings for the lab-frame parameters used during our parameter estimation. The definitions of the parameters are documented in \citet[Table~E1]{bilby_gwtc}. The trigger time $t_c$ is obtained from the data products of \pycbc's O2 search.  \label{tab:priors}} 
    \begin{tabular}{c c c c}
    \hline
    Parameter & Shape & Limits \\
    \hline
          $\mathcal{M}\ [\msun]$           & Uniform & 7--180  \\
          $q$                           & Uniform & 0.1--1  \\
          $M\ [\msun]$                     & Constraint & 50--500  \\
          $d_\mathrm{L}\ [\mathrm{Mpc}]$   & Comoving & 100--5000  \\
          $\chi_1$, $\chi_2$            & Uniform & -1--1  \\
          $\theta_{JN}$                 & Sinusoidal & 0--$\pi$  \\
          $\psi$                        & Uniform & 0--$\pi$  \\
          $\phi$                        & Uniform & 0--$2\pi$  \\
          ra                            & Uniform & 0--$2\pi$  \\
          dec                           & Cosine & 0--$2\pi$  \\
          $t_c\ [\mathrm{s}]$              & Uniform & $t_c\pm0.1$  \\
    \hline
    \end{tabular}
\end{table}

The waveform template we utilize is \imrphenomp, a phenomenological waveform template constructed in the frequency domain that models the in-spiral, merger, and ring-down (IMR) of a compact binary coalescence~\citep{khan2016frequency}. Although there exist gravitational-wave templates such as \imrxhm~\citep{imrphenompxhm}, \nrsur~\citep{nrsur7dq4} and \seob~\citep{seobnrv4phm} which incorporate more physics, such as information on higher-order modes, we use \imrphenomp as it is computationally inexpensive compared to others.

To generate the PSD, we take 31 neighboring off-source non-overlapping  4-second  segments of time-series data before the analysis data segment $d_i$. A Tukey window with a 0.2-second roll-off is applied to each data segment to suppress spectral leakage. After this, we fast-Fourier transform and median-average the segments to create a PSD~\citep{ligo_psd}. Like other PSD estimation methods, this method adds statistical uncertainties to the PSD~\citep{psd_student_t, chatziioannou2019noise, Biscoveanu:2020:PhRvD}. To marginalize over the statistical uncertainty, we use the median-likelihood presented by~\citet{psd_student_t} as a post-processing step. 
This post-processing step reduces the percentage of background $\bgrdbcr>0$ by $\sim49\%$.
The details of this calculation are in the Appendix~\ref{sec:psd-marginalization}.

The data we use are the publicly accessible O2 strain data from the Hanford and Livingston detectors, recorded while the detectors are in ``Science Mode''. We obtain the data from the gravitational-wave Open Science Center~\citep{GWOSC} using \gwpy~\citep{gwpy}. 

Finally, with the $\candbcr$ and $\bgrdbcr$ for each time-frame of triggers, we calculate the candidate signal's $\pastrobcr$. 

\section{Results}
\label{sec:Results}

We analyze the O2 candidates with ${\totMlab > 55~\msun}$ and report candidates with ${\pastrobcr\geq0.2}$ in Table~\ref{tab:results}. The $\hat{\pi}^S$ and $\hat{\pi}^G$ values utilized for each time-frame are reported in Appendix~\ref{apdx:alphabeta}. By imposing a $\pastrobcr$ threshold of 0.5, we present 13 candidate gravitational wave events. 

\begin{table*}
\centering
\caption{
$\pastrobcr$ table for gravitational wave events and candidates in our search space with $\pastrobcr>0.2$, calculated using Hanford and Livingston observatory data. Displayed for comparison are significances of events taken from: GstLAL \pastroGwtcGstlal~\citep{GWTC1}, 
PyCBC \pastroGwtcPycbc~\citep{GWTC1}, 
IAS \pastroIas~\citep{IAS1, IAS2},  
\pastroPrat~\citep{bayesian_odds}, 
PyCBC OGC-2 \pastroOgcTwo~\citep{pycbc_ogc_2} and
PyCBC OGC-3 \pastroOgcThree~\citep{pycbc_ogc_2}. 
The $\tc$ column contains the `GPS' coalescence-times of the gravitational wave events. 
The catalog column displays the first catalog reporting the event on each row (the catalogs labeled IAS-1 and IAS-2 correspond to the candidates published by \citealt{IAS1} and \citealt{IAS2}).
}
\label{tab:results}

\def\arraystretch{1.5} 
\setlength{\tabcolsep}{0.5em}
\begin{NiceTabular}{@{}ll!{\quad}|c|cc!{\quad}c!{\quad}c!{\quad}cc!{\quad}|c@{}}
\CodeBefore
\rowcolors{2}{white}{gray!10}
\Body

     Event & Catalog &  \pastrobcr &  \pastroGwtcPycbc &  \pastroGwtcGstlal &  \pastroIas &  \pastroPrat &    \pastroOgcTwo &  \pastroOgcThree &          \tc \\
\hline
  GW170104 &  GWTC-1 &        0.97 &              1.00 &               1.00 &             &         1.00 &             1.00 &                  & 1167559936.60 \\
  GW170121 &   IAS-1 &        0.83 &                   &                    &        1.00 &         0.53 &             1.00 &             1.00 & 1169069154.57 \\
    170209 &       - &        0.32 &                   &                    &             &              &                  &                  & 1170659643.47 \\
    170222 &       - &        0.58 &                   &                    &             &              &                  &                  & 1171814476.97 \\
    170302 &   IAS-1 &        0.78 &                   &                    &        0.45 &              &                  &                  & 1172487817.48 \\
  GW170304 &   IAS-1 &        0.94 &                   &                    &        0.99 &         0.03 &             0.70 &             0.70 & 1172680691.36 \\
 GWC170402 &   IAS-2 &        0.60 &                   &                    &        0.68 &         0.00 &                  &                  & 1175205128.57 \\
  GW170403 &   IAS-1 &        0.54 &                   &                    &        0.56 &         0.27 &             0.03 &             0.71 & 1175295989.22 \\
    170421 &       - &        0.27 &                   &                    &             &              &                  &                  & 1176789158.14 \\
  GW170425 &   IAS-1 &        0.22 &                   &                    &        0.77 &         0.74 &             0.21 &             0.41 & 1177134832.18 \\
  GW170608 &  GWTC-1 &        0.99 &              1.00 &               0.92 &             &         1.00 &                  &                  & 1180922494.50 \\
  GW170727 &   IAS-1 &        0.98 &                   &                    &        0.98 &         0.66 &             0.99 &             1.00 & 1185152688.02 \\
  GW170729 &  GWTC-1 &        0.98 &              0.52 &               0.98 &             &         1.00 &             1.00 &             0.99 & 1185389807.30 \\
  GW170809 &  GWTC-1 &        0.99 &              1.00 &               0.99 &             &         1.00 &             1.00 &             1.00 & 1186302519.75 \\
  GW170814 &  GWTC-1 &        1.00 &              1.00 &               1.00 &             &         1.00 &             1.00 &             1.00 & 1186741861.53 \\
 GW170817A &   IAS-2 &        0.92 &                   &                    &        0.86 &         0.02 &                  &                  & 1186974184.72 \\

\end{NiceTabular}
\end{table*}

Various search pipeline \pastro are not mathematically equivalent~\citep{Galaudage:2020:PhRvD}. Moreover, \pastro is not equivalent to $\pastrobcr$. However, by comparing candidates' various \pastro values with $\pastrobcr$, we can compare how significant each pipeline deems the candidate. For comparison, in Table~\ref{tab:results} we report \pastro values from \GWTC~\citep{GWTC1}, PyCBC OGC-2~\citep{pycbc_ogc_2}, PyCBC OGC-3~\citep{pycbc_ogc_2}, IAS~\citep{IAS1, IAS2}, and \citet{bayesian_odds}'s analyses.

\subsection{GWTC-1 Events}
All the confirmed gravitational-wave events from binary black hole mergers reported in \GWTC and within our prior space (specifically GW170104, GW170608, GW170729, GW170809, and GW170814) have ${\pastrobcr>0.9}$, indicating a high probability of an astrophysical signal. 

In addition to the above confirmed gravitational-wave events from \GWTC, we have also analyzed several candidate events from \GWTC, most of which have low $\pastrobcr$. For example, consider the candidate event 170412 (${\tc = 1176047817}$), assigned a $\pastro$ of $0.06$ by \gstlal and has a $\pastrobcr$ of $0.01$. This candidate was reported to be excess power caused due to noise appearing non-stationary between ${60-200\ \text{Hz}}$~\citep{GWTC1}. This candidate demonstrates that $\pastrobcr$ may be utilized to eliminate candidates originating from terrestrial noise sources. 

\subsection{IAS Events}
Our analysis of the IAS events and candidates with $\totMlab\gtrsim55~\msun$ in O2 has resulted in one event with disfavored ${\pastrobcr<0.5}$ (GW170425), and five events and two candidates with $\pastrobcr\geq0.5$ (GW170121, GW170304, 170302, GWC170402, GW170403, GW170727, GW170817A). From this list, four events (GW170121, GW170304, GW170727, GW170817A) have ${\pastrobcr>0.8}$ and ${\pastro>0.9}$ reported from other pipelines, making them viable gravitational-wave event candidates.  

GWC170402, detected by \citet{IAS2}, is reported to originate from a binary with non-zero eccentricity~\citep{IAS2}. As we used a non-eccentric waveform during analysis, we may be under estimating this event's significance at ${\pastrobcr\leq0.6}$. Finally, GW170425 which has ${\pastrobcr<0.25}$ also has low \pastro reported in OGC-2 and OGC-3 \citep{pycbc_ogc_2,pycbc_ogc_3}, suggesting that GW170425 may have been false alarm.

\subsection{New Candidate Events}
Although no IMBH detections are made with the \bcr, a marginal stellar mass black hole merger candidate 170222 has been discovered with a ${\pastrobcr\sim0.6}$. This candidate has a ${\text{SNR}\sim7.7}$, low spin magnitudes, and source-frame component masses of ${({47.16}_{-5.77}^{+8.00}, {35.50}_{-6.35}^{+5.79}) \msun}$ ($90\%$ credible intervals), making it one of the heavier black-hole mergers from O2 and \GWTC. This candidate may be of interest as one component black hole may lie in the pair-instability mass gap $({55^{+10}_{-10}-148^{+13}_{-12})\msun}$~\citep{Woosley:2021:arXiv, Heger:2002:ApJ}. More details on the candidate are presented in Appendix~\ref{apdx:170222}. The remaining coherent trigger candidates all have ${\pastrobcr<0.5}$, making them unlikely to originate from astrophysical sources. 

\section{Conclusion}
\label{sec:Conclusion}

In this paper, we demonstrate that the Bayesian Coherence Ratio \bcr~\citep{BCR1} can be used as a ranking statistic to provide a measure of significance for gravitational-wave signals originating from CBCs with lab-frame total masses between $55~\msun$ and $400~\msun$, a range that includes IMBHs. To compute the \bcr for candidates, we utilize Bayesian inference to calculate the probability of data under various hypotheses (the hypotheses that the data contains a coherent signal, just noise, or an incoherent glitch). This Bayesian ranking method takes a step towards building a unified Bayesian framework that provides a measure of significance for candidates and estimates their parameters, utilizing the same level of physical information incorporated during detected parameter estimation studies. 

In our study, we analyze O2 binary-black hole events and candidates with ${\totMlab > 55~\msun}$ reported by the \pycbc search~\citep{pycbc_ogc_2}, the \IAS-team~\citep{IAS1, IAS2} and those reported in \GWTC~\citep{GWTC1}. Using a $\pastrobcr$ threshold of 0.5, we find that the \GWTC events have high probabilities of originating from an astrophysical source. We also find that some of the \GWTC marginal triggers that have corroborated terrestrial sources (for example, candidate 170412) have low $\pastrobcr$, indicating this method's ability to discriminate between terrestrial artifacts and astrophysical signals. Our analysis of the \IAS events demonstrates that GW170121, GW170304, GW170727, and GW170817A are likely to originate from astrophysical sources (${\pastrobcr\geq0.8}$), while GW170425 is not (${\pastrobcr<0.25}$). Finally, we report a new marginal binary-black hole merger candidate, 170222. 

With the rapid rate of development in gravitational-wave Bayesian inference, we anticipate the ability to analyze longer-duration signals, utilize more advanced signal and glitch models, and incorporate data from the entire detector network. In a similar vein, with the accumulation of more gravitational wave events, future \bcr work may utilize astrophysically informed priors during Bayesian inference and more accurate prior probabilities for each detector.

\section*{Acknowledgments}{
The authors gratefully thank the \pycbc team for providing the gravitational-wave foreground, background, and simulated triggers from \pycbc's search of O2's data. We also warmly thank Ian Harry and Thomas Dent for answering questions about the \pycbc search's data products.  

We gratefully acknowledge the computational resources provided by the LIGO Laboratory—Caltech Computing Cluster and supported by NSF grants PHY-0757058 and PHY-0823459, and thank Stuart Anderson for his assistance in resource scheduling.

All analyses (inclusive of test and failed analyses) performed for this study used ${0.6\mathrm{M}}$ core-hours, amounting to a carbon footprint of ${\sim77\ \mathrm{t}}$ of ${\text{CO}_2}$ (using the U.S. average electricity source emissions of ${0.371\ \text{kg/kWh}}$~\citep{greenhouse} and ${0.3\ \text{kWh}}$ for each CPU).

This material is based upon work supported by NSF’s LIGO Laboratory, a major facility fully funded by the National Science Foundation. This research has used data, software, and web tools obtained from the Gravitational Wave Open Science Center (\href{https://www.gw-openscience.org}{https://www.gw-openscience.org}), a service of LIGO Laboratory, the LIGO Scientific Collaboration and the Virgo Collaboration. LIGO Laboratory and Advanced LIGO are funded by the United States National Science Foundation (NSF) as well as the Science and Technology Facilities Council (STFC) of the United Kingdom, the Max-Planck-Society (MPS), and the State of Niedersachsen/Germany for support of the construction of Advanced LIGO and construction and operation of the GEO600 detector. Additional support for Advanced LIGO was provided by the Australian Research Council. Virgo is funded, through the European Gravitational Observatory (EGO), by the French Centre National de Recherche Scientifique (CNRS), the Italian Istituto Nazionale di Fisica Nucleare (INFN) and the Dutch Nikhef, with contributions by institutions from Belgium, Germany, Greece, Hungary, Ireland, Japan, Monaco, Poland, Portugal, Spain.

}

\section*{Data Availability}
{
We analyze publicly-available gravitational wave strain data from the LIGO-Virgo-KAGRA collaboration~\citep{dataO2}. The trigger times for analysis were provided by the \pycbc team~\citep{pycbc_ogc_2}. The derived data generated in this research will be shared on reasonable request to the corresponding author.
\begin{flushleft}
\textit{Software:} \href{https://lscsoft.docs.ligo.org/bilby/}{\bilby}~\citep[v0.6.8]{bilby}, \href{https://lscsoft.docs.ligo.org/bilby_pipe/master/index.html}{\bilbypipe}~\citep[v0.3.12]{bilby_pipe}, \href{https://dynesty.readthedocs.io/}{\dynesty}~\citep[v0.9.5.3]{dynesty}, \href{https://gwpy.github.io/docs/stable/index.html}{\gwpy}~\citep[v1.0.1]{gwpy}, \href{https://lscsoft.docs.ligo.org/lalsuite/lalsimulation/index.html}{\code{LALSimulation}}~\citep[v6.70]{lalsuite}, \code{matplotlib}~\citep[v3.2.0]{matplotlib}, \code{NumPy}~\citep[v1.8.1]{NumPy}, \code{SciPy}~\citep[v1.4.1]{SciPy}, \code{pandas}~\citep[v1.0.2]{pandas}, \code{python}~\citep[v3.7]{pythonForScientificComputing,pythonForScientists}. 
\end{flushleft}
}



\bibliographystyle{mnras}
\bibliography{high_mass_bib} 


\appendix

\section{Bayesian Evidence Evaluation}\label{sec:bayesianEvidEval}

\subsection{Noise Model}
We assume that each detector's noise is Gaussian and stationary over the period being analyzed~\citep{ligo_psd}. In practice, we assume that the noise has a mean of zero that the noise variance $\sigma^2$ is proportional to the noise power spectral density \psd of the data. Using \psd, for each frequency-domain data segment $d_i$ in each of the $i$ detectors in a network of $D$ detectors, we can write 
\begin{equation}
\label{eq:zn}
Z^N_i = \mathcal{N}(d_i|\mu=0,\sigma^2=\psd),
\end{equation}
where $\mathcal{N}$ is a normal distribution. 

\subsection{Coherent Signal Model}
We model coherent signals using a binary black hole waveform template \template, where the vector \parameters contains a point in the 11-dimensional space describing aligned-spin binary-black hole mergers. For the signal to be coherent, \parameters must be consistent in each 4-second data segment $d_i$ for a network of $D$ detectors. Hence, the coherent signal evidence is calculated as
\begin{equation}
\label{eq:zs}
Z^S = \int\limits_{\parameters} \prod\limits^{D}_{i=1} \left[ \mathcal{L}(d_i|\template)\right] \pi(\parameters | \mathcal{H}_S)\  \text{d}\parameters \ ,
\end{equation}
where $\pi(\parameters| \mathcal{H}_S)$ is the prior for the parameters in the coherent signal hypothesis $\mathcal{H}_S$, and $\mathcal{L}(d_i|\template)$ is the likelihood for the coherent signal hypothesis that depends on the gravitational-wave template \template and its parameters \parameters. 

\subsection{Incoherent Glitch Model}
Finally, as glitches are challenging to model and poorly understood, we follow \citet{bci} and utilize a surrogate model for glitches. The glitches are modeled using gravitational-wave templates  \template with uncorrelated parameters amongst the different detectors such that  $\parameters_i \neq \parameters_j$ for two detectors $i$ and $j$~\citep{bci}.  Modeling glitches with \template captures the worst-case scenario: when glitches are identical to gravitational-wave signals (excluding coherent signals). Thus, we can write $Z^G_i$ as 
\begin{equation}
\label{eq:zg}
Z^G_i = \int\limits_{\parameters} \mathcal{L}(d_i|\template)\ \pi(\parameters| \mathcal{H}_G)\  \text{d}\parameters  \ ,
\end{equation}
where $\pi(\theta| \mathcal{H}_G)$ is the prior for the parameters in the incoherent glitch hypothesis $\mathcal{H}_G$.

\section{Tuning the prior probabilities}\label{apdx:tuning-prior-odds}

After calculating the \bcr for a set of background triggers and simulated triggers from a stretch of detector-data (a data chunk), we can compute probability distributions for the background and simulated triggers, $p_b(\bcr)$ and $p_s(\bcr)$. We expect the background trigger and simulated signal \bcr values to favor the incoherent glitch and the coherent signal hypothesis, respectively. Ideally, these distributions representing two unique populations should be distinctly separate and have no overlap in their \bcr values. The prior probability parameters $\hat{\pi}^S$ and $\hat{\pi}^G$ from Eq.~\ref{eq:bcr} help separate the two distributions. Altering $\hat{\pi}^S$ translates the \bcr probability distributions while adjusting $\hat{\pi}^G$ spreads the distributions~\citep[see][Appendix A]{BCR1}. Although Bayesian hyper-parameter estimation can determine the optimal values for $\hat{\pi}^S$ and $\hat{\pi}^G$, an easier approach is to adjust the parameters for each data chunk's \bcr distribution. In this study, we tune $\hat{\pi}^S$ and $\hat{\pi}^G$ to maximally separate the \bcr distributions for the background and simulated triggers. 

To calculate the separation between $p_b(\bcr)$ and $p_s(\bcr)$, we use the Kullback--Leibler divergence (KL divergence) $D_{KL}$, given by
\begin{equation}
    D_{KL}(p_b | p_s) = \sum\limits_{x\in \bcr} p_b(x) \log \left( \frac{p_b(x)}{p_s(x)} \right)  \ .
\end{equation}
The $D_{K.L.}=0$ when the distributions are identical and increases as the asymmetry between the distributions increases. 

We limit our search for the maximum KL-divergence in the $\hat{\pi}^S$ and $\hat{\pi}^G$ ranges of $[10^{-10}, 10^0]$. We set our values for $\hat{\pi}^S$ and $\hat{\pi}^G$ to those which provide the highest KL-divergence and calculate the \bcr for candidate events present in this data chunk. Note that we conduct the analysis in data chunks of two weeks rather than an entire data set of a few months as the background may be different at different points of the entire data set.

\section{Marginalizing over PSD statistical uncertainties}\label{sec:psd-marginalization}
\begin{figure*}
    \centering
    \begin{subfigure}
        \centering
        \includegraphics[width=0.45\linewidth]{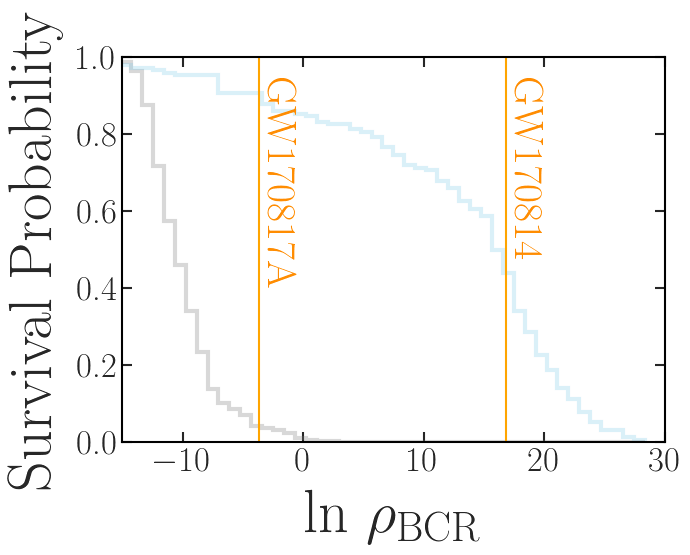}
    \end{subfigure}
    ~ 
    \begin{subfigure}
        \centering
        \includegraphics[width=0.45\linewidth]{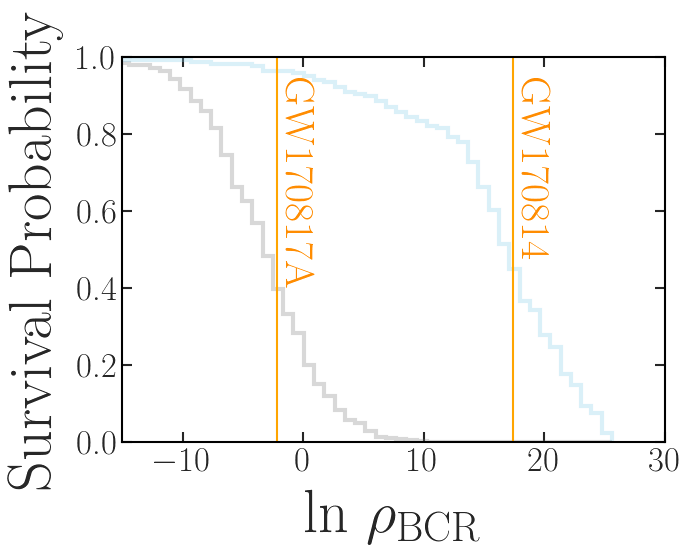}
    \end{subfigure}
    \caption{
    Histograms represent the survival function (1-CDF) from our selection of background triggers (gray) and simulated signals (blue) triggers obtained from \pycbc's search of data from $\text{August 13 - 21, 2017}$. Vertical lines mark the $\text{ln}\ \bcr$ of \IAS's GW170817A and \GWTC's GW170814.
    Left: Survival functions using the post-processing step to marginalize over PSD statistical uncertainties. Right: Survival functions without the post-processing step. Without the post-processing step, there is a greater overlap between the background (gray) and foreground (blue) survival functions.
    \label{fig:bcrCdf}}
\end{figure*}

To generate the results presented in Table~\ref{tab:results}, we applied a post-processing step to marginalize the uncertainty in the PSD. In Fig.~\ref{fig:bcrCdf}, we demonstrate the impact of the post-processing step. Marginalizing over uncertainty in the PSD yields an improvement in the separation of the noise and signal distributions (left plot). Quantitatively, at a threshold $\bcr^T=0$ the post-processing step reduces the percentage of background $\bcr > \bcr^T$ from $60\%$ to $25\%$ (a $58\%$ improvement) in the August 13 - 21, 2017 time-frame of data. For the entirety of O2, PSD marginalization reduces the percentage of $\bcr > \bcr^T$ from $64\%$ to $33\%$ (a $\sim49\%$ improvement). 

\section{Tuned prior probabilities}\label{apdx:alphabeta}

O2 lasted several months, over which the detector's sensitivity varied. Hence, a part of our analysis entailed tuning the prior probabilities for obtaining a signal and a glitch, $\hat{\pi}^S$ and $\hat{\pi}^G$, as described in Section~\ref{sec:method}. Table~\ref{tab:priorodds} presents the signal and glitch prior probabilities utilized for each time-frame of O2 data. 

Tuning the prior probabilities can dramatically affect the $\pastrobcr$. For example, consider Table~\ref{tab:tuningresults}, which reports tuned $\pastrobcr$ and un-tuned \untunedpastrobcr (where $\hat{\pi}^S=1$ and $\hat{\pi}^G=1$) for various events and candidates. By tuning the prior probabilities, the $\pastrobcr$ for some IAS events (for example, GW170403 and GW170817A) can change by more than 0.5, resulting in the promotion/demotion of a candidate's significance.

\begin{table}
\centering
\caption{The prior odds used for each time-frame of data from O2. Each time frame commences at the start date and concludes at the following time-frame's start date.
    }
\label{tab:priorodds}
\def\arraystretch{1.5} 
\setlength{\tabcolsep}{0.5em}
\begin{NiceTabular}{c|cc}
\CodeBefore
\rowcolors{2}{white}{gray!10}
\Body

 Start Date &    $\hat{\pi}^S$ &    $\hat{\pi}^G$ \\
\midrule
 2016-12-23 & 1.00E+00 & 6.25E-01 \\
 2017-01-22 & 1.00E+00 & 2.33E-02 \\
 2017-02-03 & 1.00E-10 & 2.44E-01 \\
 2017-02-12 & 1.76E-08 & 5.96E-02 \\
 2017-02-20 & 6.55E-10 & 2.22E-03 \\
 2017-02-28 & 1.00E-10 & 5.96E-02 \\
 2017-03-10 & 2.56E-10 & 3.91E-01 \\
 2017-03-18 & 1.60E-10 & 1.00E+00 \\
 2017-03-27 & 1.10E-08 & 5.96E-02 \\
 2017-04-04 & 3.73E-02 & 2.33E-02 \\
 2017-04-14 & 1.05E-09 & 2.44E-01 \\
 2017-04-23 & 2.68E-09 & 1.46E-02 \\
 2017-05-08 & 1.00E+00 & 2.44E-01 \\
 2017-06-18 & 6.55E-10 & 3.39E-04 \\
 2017-06-30 & 2.02E-05 & 5.69E-03 \\
 2017-07-15 & 1.05E-09 & 9.54E-02 \\
 2017-07-27 & 1.00E+00 & 2.12E-04 \\
 2017-08-05 & 2.12E-04 & 3.73E-02 \\
 2017-08-13 & 2.68E-09 & 8.69E-04 \\

\end{NiceTabular}
\end{table}

\begin{table}
\centering
\caption{Table of \pastrobcr using ``tuned'' prior odds and  \pastrobcr using uninformed prior odds of 
$\hat{\pi}^S=1$ and $\hat{\pi}^G=1$ (represented by \untunedpastrobcr).  
Details of other columns provided in Table~\ref{tab:results}.}
\label{tab:tuningresults}

\def\arraystretch{1.5} 
\setlength{\tabcolsep}{0.5em}
\begin{NiceTabular}{ll|c c| c}
\CodeBefore
\rowcolors{2}{white}{gray!10}
\Body
     Event & Catalog & \tunedpastrobcr & \untunedpastrobcr &          \tc \\
\hline
  GW170104 &  GWTC-1 &             0.97 &               0.95 & 1167559936.60 \\
  GW170121 &   IAS-1 &             0.83 &               0.68 & 1169069154.57 \\
    170209 &       - &             0.32 &               0.00 & 1170659643.47 \\
    170222 &       - &             0.58 &               0.50 & 1171814476.97 \\
    170302 &   IAS-1 &             0.78 &               0.54 & 1172487817.48 \\
  GW170304 &   IAS-1 &             0.94 &               0.80 & 1172680691.36 \\
 GWC170402 &   IAS-2 &             0.60 &               0.00 & 1175205128.57 \\
  GW170403 &   IAS-1 &             0.54 &               0.90 & 1175295989.22 \\
    170421 &       - &             0.27 &               0.21 & 1176789158.14 \\
  GW170425 &   IAS-1 &             0.22 &               0.16 & 1177134832.18 \\
  GW170608 &  GWTC-1 &             0.99 &               0.99 & 1180922494.50 \\
  GW170727 &   IAS-1 &             0.98 &               0.99 & 1185152688.02 \\
  GW170729 &  GWTC-1 &             0.98 &               0.95 & 1185389807.30 \\
  GW170809 &  GWTC-1 &             0.99 &               0.99 & 1186302519.75 \\
  GW170814 &  GWTC-1 &             1.00 &               1.00 & 1186741861.53 \\
 GW170817A &   IAS-2 &             0.92 &               0.30 & 1186974184.72 \\

\end{NiceTabular}
\end{table}

\section{A closer look at 170222}\label{apdx:170222}
\begin{figure*}
    \centering
    \begin{subfigure}
        \centering
        \includegraphics[width=0.45\linewidth]{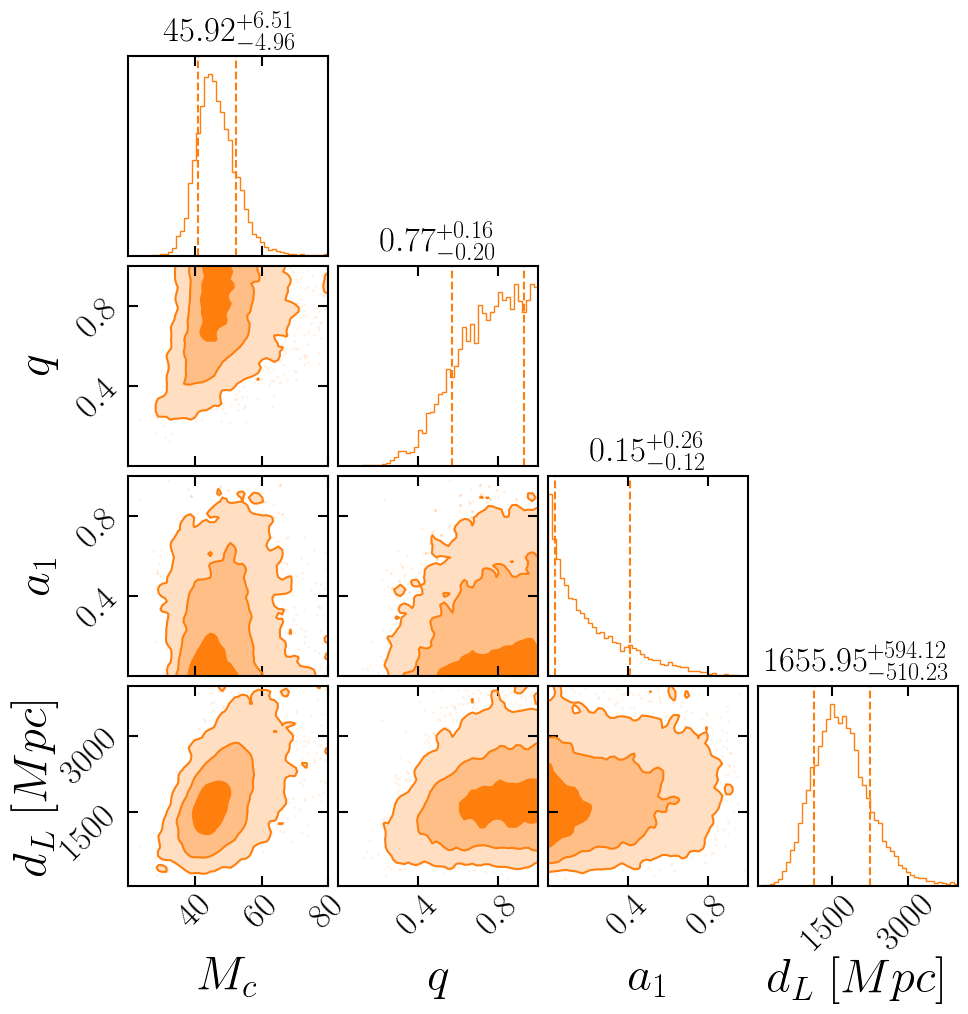}
    \end{subfigure}
    ~ 
    \begin{subfigure}
        \centering
        \includegraphics[width=0.45\linewidth]{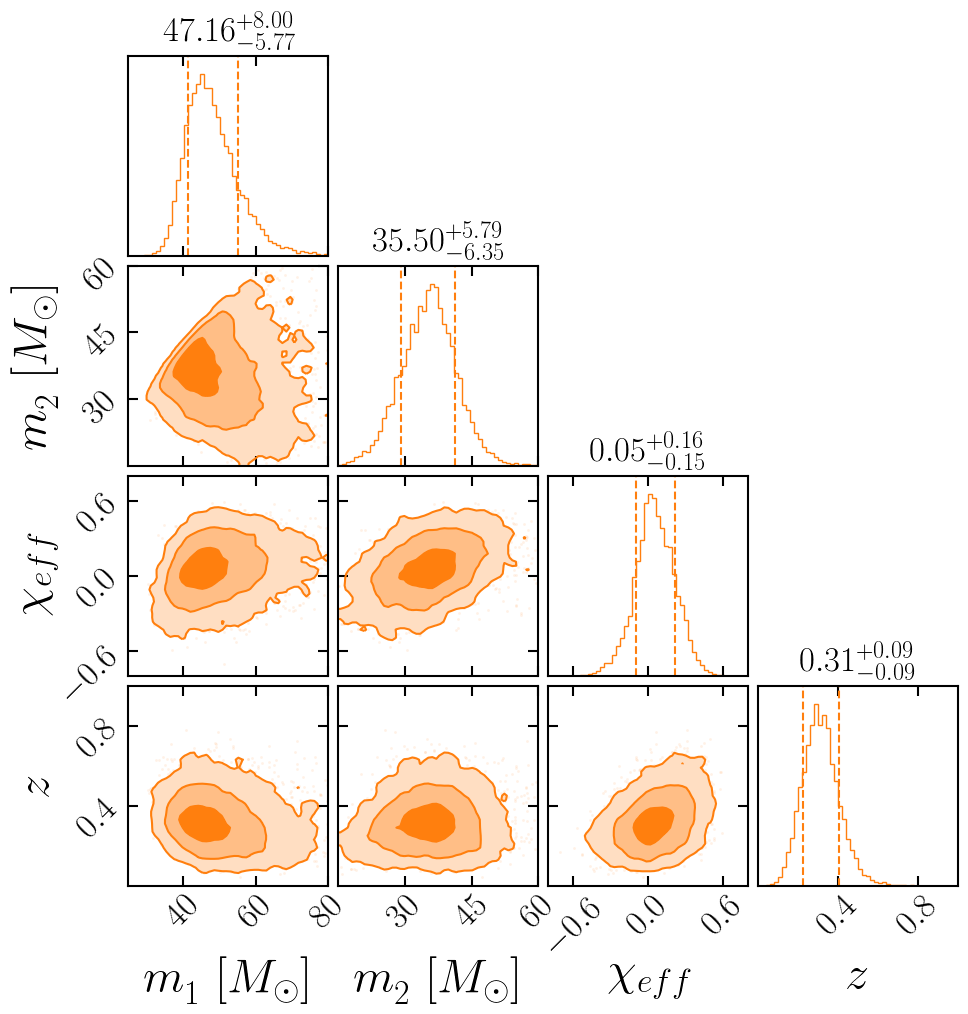}
    \end{subfigure}
    \caption{Posterior distributions for 8 parameters of 170222. 
    Left: Posterior probability distributions for 4 of the 12 search parameters.
    Right: Posterior probability distributions for 4 derived parameters.
    \label{fig:170222}}
\end{figure*}

PyCBC found the candidate 170222 with $\mathcal{M}=49.46\ \msun$ and $q=0.68$, values contained inside the $90\%$ credible intervals of our posterior probability distributions for 170222. Some of the posteriors produced as a by-product of our \bcr calculation can be viewed in Fig.~\ref{fig:170222}.

\bsp	
\label{lastpage}

\end{document}